\documentclass[reprint,secnumarabic,amssymb, nobibnotes, aps, prl,groupedaddress,superscriptaddress,raggedbottom]{revtex4-2}
\usepackage{graphicx}
\usepackage{amsmath}
\usepackage{xcolor}
\usepackage{soul}
\usepackage[normalem]{ulem}

\definecolor{midnightblue}{rgb}{0.098,0.098,0.439}

\begin{document}
	\title{Anisotropic heating and magnetic field generation due to Raman scattering in laser-plasma interactions}%
	\author{T. Silva}%
	\email{thales.silva@tecnico.ulisboa.pt}
	\affiliation{GoLP/Instituto de Plasmas e Fus\~ao Nuclear, Instituto Superior T\'ecnico, Universidade de Lisboa, 1049-001 Lisboa, Portugal}
	\author{K. Schoeffler}%
	\affiliation{GoLP/Instituto de Plasmas e Fus\~ao Nuclear, Instituto Superior T\'ecnico, Universidade de Lisboa, 1049-001 Lisboa, Portugal}
	\author{J. Vieira}%
	\affiliation{GoLP/Instituto de Plasmas e Fus\~ao Nuclear, Instituto Superior T\'ecnico, Universidade de Lisboa, 1049-001 Lisboa, Portugal}
	\author{M. Hoshino}%
	\affiliation{Department of Earth and Planetary Science, University of Tokyo, Tokyo 113-0033, Japan}
	\author{R. A. Fonseca}%
	\affiliation{GoLP/Instituto de Plasmas e Fus\~ao Nuclear, Instituto Superior T\'ecnico, Universidade de Lisboa, 1049-001 Lisboa, Portugal}
	\affiliation{DCTI/ISCTE Lisbon University Institute, 1649-026 Lisbon, Portugal}
	\author{L. O. Silva}%
	\email{luis.silva@tecnico.ulisboa.pt}
	\affiliation{GoLP/Instituto de Plasmas e Fus\~ao Nuclear, Instituto Superior T\'ecnico, Universidade de Lisboa, 1049-001 Lisboa, Portugal}
	\date{\today}%
	
	\begin{abstract}
	We identify a novel mechanism for magnetic field generation in the interaction of intense electromagnetic waves and underdense plasmas. We show that Raman scattered plasma waves trap and heat the electrons preferentially in their propagation direction, resulting in a temperature anisotropy. In the trail of laser pulse, we observe magnetic field growth which matches the Weibel mechanism due to the temperature anisotropy. We discuss the role of the initial electron temperature in our results. The predictions are confirmed with multi-dimensional particle-in-cell simulations. We show how this configuration is an experimental platform to study the long-time evolution of the Weibel instability.
	\end{abstract}
	\maketitle
	
	\section{Introduction}
	Understanding the interaction between electromagnetic waves and plasmas is of fundamental importance with implications in inertial confinement fusion, plasma-based accelerators, and laboratory astrophysics. The ability to use the interaction of intense lasers with plasmas to reproduce astrophysical scenarios in the laboratory is a powerful tool to explore astrophysical phenomena \cite{1999Remington,2019Zhang}, specifically in connection with laser-plasma produced magnetic fields. Several mechanisms for magnetic field generation, such as the Biermann battery and the inverse Faraday effect, have been discussed over the years \cite{1975Stamper,1978Max,1979Raven,1991Stamper,1994Lehner,1994Askaryan,1996Pukhov,1997Pegoraro,2015Marocchino}. Laser-plasma interactions can generate strong magnetic fields, as shown in experimental and numerical studies of magnetic turbulence \cite{2017Chatterjee}, laser wakefield acceleration \cite{2015Flacco}, and collisionless shocks \cite{2012Fiuza,2013Fox,2015Huntington}.
	\label{key}
	In the context of intense electromagnetic wave interaction with underdense plasmas, effects such as stimulated Raman scattering (SRS) are important, and can determine the magnetic field generation process, as electrons can heat and drive strong currents in the plasma \cite{1985Forslund,1986Mori,2010Masson}. Stimulated Raman scattering is the resonant decay of a photon into another photon and an electron plasma wave. Theory and growth rates for SRS were studied extensively in recent decades \cite{1974Drake,1983Estabrook,1988McKinstrie,1992McKinstrie,1992Antonsen,1992Wilks,1993Antonsen,1994Sakharov}. This phenomena occurs only for underdense plasmas with density $n \lesssim n_{cr}/4$ ($n_{cr}=m_e\omega_0^2/4\pi e^2$ is the critical density, $e$ is the elementary charge, $m_e$ the electron mass, and $\omega_0$ is the pump frequency). Energy and momentum conservation require that the frequencies and wavevectors obey
	 \begin{subequations}\label{match}
		\begin{gather}
		\omega_0 = \omega_{pw} + \omega_{s},\label{frq_match}\\
		\mathbf{k}_0 = \mathbf{k}_{pw} + \mathbf{k}_{s},\label{wv_match}
		\end{gather}
	\end{subequations}
	where the subscripts ``0'', ``pw'', and ``s'' refer to the pump, and scattered plasma and scattered electromagnetic waves, respectively. Much of the attention drawn to SRS is related to inertial fusion, where the plasma waves associated with the SRS instability can preheat the plasma electrons \cite{1980Estabrook,1982Phillion,1984Figueroa,2018Rosenberg}. Furthermore, there is a more recent interest in using plasmas as a medium for amplification of electromagnetic pulses through the SRS mechanism \cite{1998Shvets,1999Malkin,2000Ping,2010Trines,2017Vieux,2018Sadler}.
	
	The fields left in the plasma after the SRS process develops (and their structure and temporal evolution) is a subject much less investigated. Reference  \cite{1986Mori} found growth of magnetic fields following SRS and explored the interaction of such fields with incoming laser light. Reference \cite{2010Masson} also shows evidence of magnetic field growth due to the current of trapped electrons on the scattered plasma waves and hints of subsequent Weibel instability. Despite these early efforts, the connection between SRS and the Weibel has not been explored, nor its interplay, long-time evolution, and dependence on the physical configuration.
	
	In this work, we identify a novel mechanism for the generation of strong magnetic fields in an underdense plasma-laser interaction. We show that the SRS scattered plasma waves heat the plasma preferentially in their propagation direction. The resulting temperature anisotropy leads to the onset of the Weibel instability. We show this setup is ideal for studying the long-time evolution of the Weibel instability, addressing the fundamental question of how magnetic fields evolve from small to long-scales. This has important implications on the structure of collisionless shocks in astrophysical objects \cite{2005Medvedev,2009Keshet}. This novel configuration allows for probing such phenomena in the laboratory.

	\section{Anisotropic heating due to stimulated Raman scattering drives Weibel instability}
	We model our scenario with kinetic particle-in-cell simulations using the OSIRIS framework \cite{2002Ricardo,2013Fonseca}. The numerical parameters for these simulations are explained in the supplemental material \cite{SM}. We simulate the interaction between an electromagnetic wave, with peak normalized vector potential  $a_0 = e A_0/m_ec^2 = 0.2$, frequency $\omega_0/\omega_{p} = 10$, where $\omega_p = (4\pi e^2 n/m_e)^{1/2}$ is the electron plasma frequency, beam waist $W_0=50c/\omega_p$ (Gaussian profile), injected from the left boundary at $t=0$ along $x_1$, and linearly polarized in the $x_3$ direction, and a uniform plasma. The ions are considered immobile, but we tested realistic ion (proton) mass and found no influence in the results at the time scales shown here. Initially, we examine a case where the electron temperature is $T = 20~\text{eV}$.
	We later discuss the role of the initial temperature in our results. 
	
	In order to measure the Weibel magnetic fields, the pulse duration $\tau$ should match the time for SRS to significantly heat the plasma $\tau_H$, i.e., the scattered plasma waves reach high amplitude (several e-foldings of the instability) and break. However, $\tau_H$ should be short enough that there is no interaction between the pump and the generated fields. Furthermore, this enables measurements of the Weibel fields, as they are separated from the laser pulse. We estimate that $\tau_H$ is 
	around 20 e-foldings of the maximum growth rate for SRS backscatter in a cold plasma, i.e.,
	    \begin{equation}
		\tau_{H} \approx 20\gamma_{SRBS}^{-1} \approx 20\times \frac{4}{a_0} \frac{\left(\omega_0/\omega_p-1\right)^{1/2}}{2\omega_0/\omega_p-1}\omega_{p}^{-1}.
		\label{tau_size}
		\end{equation}
	We thus performed simulations using $\tau = \tau_{H} \approx 20\pi\omega_p^{-1}$.
	
		\begin{figure}[b]
		\centering
		\includegraphics[width=0.99\linewidth]{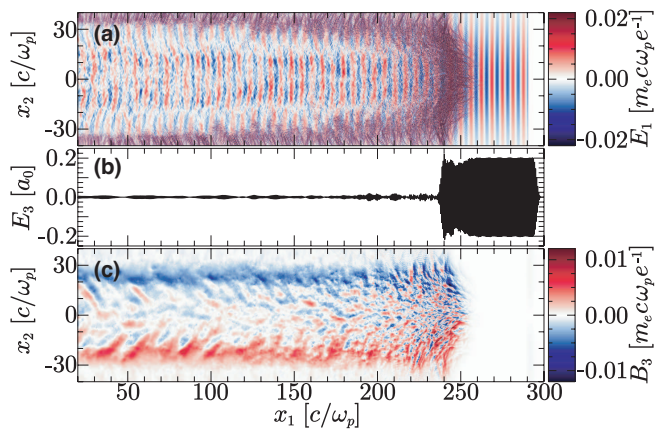}
		\caption{Illustration of the magnetic field generation process. (a) Longitudinal electric field $E_1$. (b) Lineout of $E_3$, the electric field in the laser polarization direction, taken at $x_2=0$. (c) $B_3$ magnetic field component.}
		\label{fields}
	\end{figure}
	Figure \ref{fields} presents the SRS scattered plasma waves, the laser, and the Weibel fields from our fiducial two-dimensional simulation at $t=300\omega_p^{-1}$. Figure \ref{fields}(a) shows the longitudinal electric field $E_1$ of the wakefield excited by the ponderomotive force of the laser pulse in the region $x_1 \in [250,290]\times c/\omega_p$. This is followed by small-scale structures, located around $x_1\approx 250c/\omega_p$, embedded in the rear region of the laser. These structures appear because of a combination of plasma waves originated from Raman back and side-scattering. The electric field in the laser polarization direction $E_3$, Fig. \ref{fields}(b), confirms that the structures in Fig. \ref{fields}(a) grow in the region where the laser is propagating. Behind the laser [see Fig. \ref{fields}(c)], we observe the growth of the $B_3$ component of the magnetic field due to the Weibel instability, evolving from small to long scales. Near $x_2 = \pm 30c/\omega_p$, we also observe a long-scale field, likely generated by the mechanism described in \cite{1994Askaryan}. For larger $W_0\omega_p/c$, this field should be less pronounced, making Weibel more prominent (cf. \cite{2018Shukla}).
	
\begin{figure}[t]
		\centering
		\includegraphics[width=0.99\linewidth]{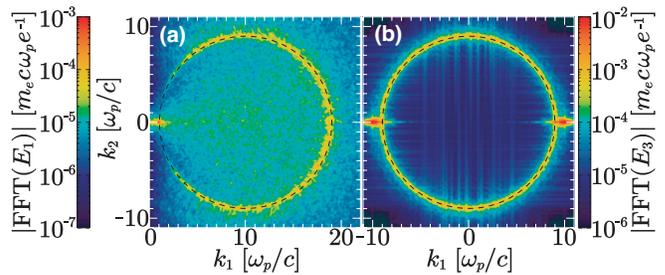}
		\caption{Fourier spectra of (a) $E_1$ and (b) $E_3$ in the region where the instability grows in Fig. \ref{fields}(a). The dashed lines are the solutions of Eqs. \eqref{matc}.}
		\label{ffte1e3}
	\end{figure}

	To confirm that the structures in Fig. \ref{fields}(a) are due to SRS and the correlation with magnetic field growth, we start by examining from Eq. \eqref{match}, the matching conditions for SRS.  Assuming a cold plasma, $\omega_0/\omega_p\gg 1$, and $\mathbf{k}_0 = k_0 \hat{x}_1$, the wavevector for the scattered plasma and electromagnetic waves respectively satisfy,
	\begin{subequations}
		\begin{gather}
		\left(k_{pw1}-k_0\right)^2 + k_{pw2}^2 = (k_0 - \omega_p/c)^2, \label{pwdisp}\\
		k_{s1}^2 + k_{s2}^2 = (k_0 - \omega_p/c)^2. \label{scdisp}
		\end{gather}
			\label{matc}
	\end{subequations}
	The Fourier spectra of $E_1$ plotted in Fig. \ref{ffte1e3}(a) reveals the dominant mode due to wakefields excited by the laser at $(k_1,k_2) = (\omega_p/c,0)$. However, since the phase velocity of this mode is close to $c$, it does not trap electrons or explain the growth of magnetic fields. Moreover, we point out the growth of a circular shape, due to the scattered plasma waves, in good agreement with Eq. \eqref{pwdisp} (dashed curve). The Fourier spectra of $E_3$ plotted in Fig. \ref{ffte1e3}(b) shows the main mode representing the incident laser field at $(|k_1|,k_2) \approx (10\omega_p/c,0)$. We also see a circular shape in good agreement with Eq. \eqref{scdisp}, thus confirming once more the presence of SRS in the system.

\begin{figure}[t]
	\centering
	\includegraphics[width=0.99\linewidth]{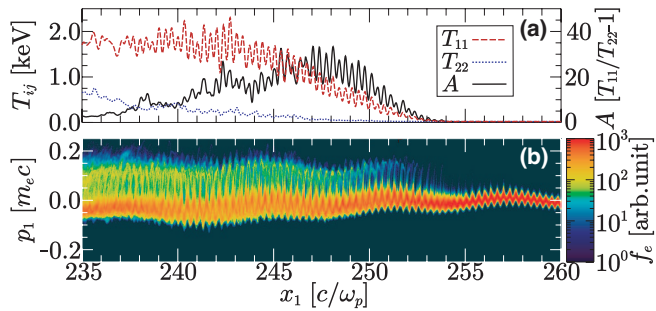}
	\caption{(a) $T_{11}$, $T_{22}$, and anisotropy parameter as a function of $x_1$. (b) Longitudinal phase space. The window is zoomed in the region of Fig. \ref{fields}(c) where we observe magnetic field growth.}
	\label{temp_wb}
\end{figure}

	 The higher intensity of the pump near the symmetry axis implies that SRS will grow predominantly in the region $|x_2|\lesssim10c/\omega_{p}$. The near backscatter modes and corresponding plasma waves stay near this region, while the side-scatter modes propagate outward. Near the axis, particle trapping in the plasma waves and wave breaking occur preferentially along $x_1$ because $k_{pw1}\gg k_{pw2}$, resulting in anisotropic heating. To confirm this, we examine the plasma temperature tensor (for $|x_2|\le5 c/\omega_{p}$) defined as
	\begin{equation}\label{def_temp}
	m_ec^2T_{ij}=\frac{\int f\left(\mathbf{p}\right) v_i p_j d^3p}{\int f\left(\mathbf{p}\right) d^3p}
	\end{equation}
	calculated in the local plasma rest frame. Figure \ref{temp_wb}(a) displays the $x_2$-averaged plasma temperature as a function of $x_1$. We also show the (averaged) anisotropy parameter defined as $A\equiv T_{\text{hot}}/T_{\text{cold}}-1$, where $T_{\text{hot}}$ and $T_{\text{cold}}$ are the larger and smaller eigenvalues of the temperature tensor $T_{ij}$, respectively. In this example, $A=T_{11}/T_{22}-1$. As predicted, $T_{11}$ and $A$ present a sharp growth at $x_1 \lesssim 254c/\omega_p$, confirming that SRS leads to an anisotropic temperature -- the driver of the Weibel mechanism \cite{1959Weibel} that amplifies small perturbations in the magnetic field. The filaments in Fig. \ref{fields}(c) are parallel to the cold direction ($x_2$ for $|x_2|\lesssim10c/\omega_p$ and oblique for $|x_2|\gtrsim10c/\omega_p$), characteristic of the Weibel instability \cite{1973Krall}, thus corroborating this mechanism for magnetic field amplification.
	 Figure \ref{temp_wb}(b) shows the longitudinal phase space density $f_e$. By comparing Figs. \ref{temp_wb}(a) and (b), it is clear that severe trapping  and wavebreaking are the main heating mechanisms as the rise in $T_{11}$ coincides with the trapping on the first plasma waves.

    \section{Role of Landau damping}
	In this example, the growth of Raman backscatter was seeded by the small thermal fluctuations $T = 20~\text{eV}$. For higher temperatures, Landau damping prevents the growth of sufficiently small wavelength plasma waves. An estimate of the wavenumber of the plasma waves for which Landau damping becomes sizable is given by $kv_{th}/\omega_p \gtrsim 0.4$ \cite{1988Kruer}, where $v_{th}/c = (T/m_ec^2)^{1/2}$ in the non-relativistic limit. As we increase the background temperature approaching this limit, the first modes that are damped are the ones near backscatter, with higher $k$. When this happens, the maximum growth rate will be at some angle between backscatter and forward scatter \cite{1992Wilks}, and thus the dominant scattered plasma waves form an angle with the pump. This affects the anisotropy direction because these plasma waves are responsible for heating the plasma preferentially in their propagation direction. We still observe magnetic field amplification behind the laser pulse, as long as the temperature is not high enough that all side-scatter modes are damped. However, because the background temperature is higher, we expect a lower anisotropy, which then leads to smaller magnetic field values.

\begin{figure}[t]
		\centering
		\includegraphics[width=0.99\linewidth]{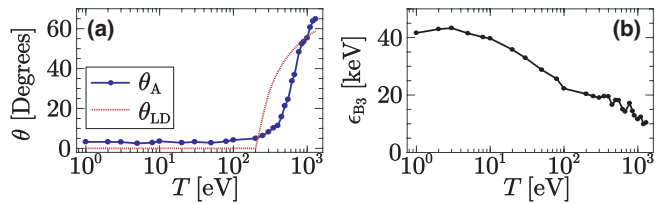}
		\caption{(a) Anisotropy angle $\theta_A$ calculated from simulation results and theoretical estimate of Landau damping $\theta_{LD}$ and (b) energy of $B_3$ as a function of the initial electron temperature.}
		\label{temp_anis}
	\end{figure}

	Landau damping plays a key role in defining the tilt angle of the anisotropy. The anisotropy direction can be found by calculating the principal axis $ \theta_A$ of the temperature tensor $T_{ij}$ \cite{2018Kevin}. Figure \ref{temp_anis}(a) shows the calculated $\theta_A$ as function of the initial electron temperature. For comparison, we also show the result of an estimate for the highest $k$ that is not Landau damped $k = 0.4\omega_p/(T/m_e)^{1/2}$ combined with Eq. \eqref{pwdisp} to calculate the angle $\theta_{LD}$ for which this effect is noticeable. By comparing $\theta_A$ and $\theta_{LD}$, we notice that the anisotropy angle remains constant while Landau damping is negligible ($T<0.1\text{ keV}$). When the temperature is enough for Landau damping to become relevant ($T>0.1\text{ keV}$), the anisotropy direction starts to change. Figure \ref{temp_anis}(b) displays the energy of the $B_3$ field, i.e., $\epsilon_{B_3}\propto\int B_3^2 d^2x$, as function of the initial plasma temperature, as observed in simulations. As predicted earlier, the field amplitude decreases for larger temperatures because the anisotropy is smaller. Nevertheless, for $T\lesssim1\text{ keV}$, field growth remains evident.
		
    \section{Seeded stimulated Raman scattering}
	To further investigate the role of SRS driven anisotropic heating, we perform additional simulations under controlled conditions (cold plasma). In addition to the driver laser, we included in our simulation seed laser pulses with $\omega_0/\omega_p=9$, $a_0 = 0.001$, and pulse duration $6\omega_p^{-1}$, with the remaining parameters the same as for the pump pulse. The seed electromagnetic wave is injected from the right boundary or from the top boundary. For these parameters, the pump and the seed waves respect the matching conditions from Eq. \eqref{frq_match}, and seed Raman backscattering and side-scattering at 90 degrees in the region where the two pulses overlap.
		\begin{figure}[t]
		\centering
		\includegraphics[width=0.99\linewidth]{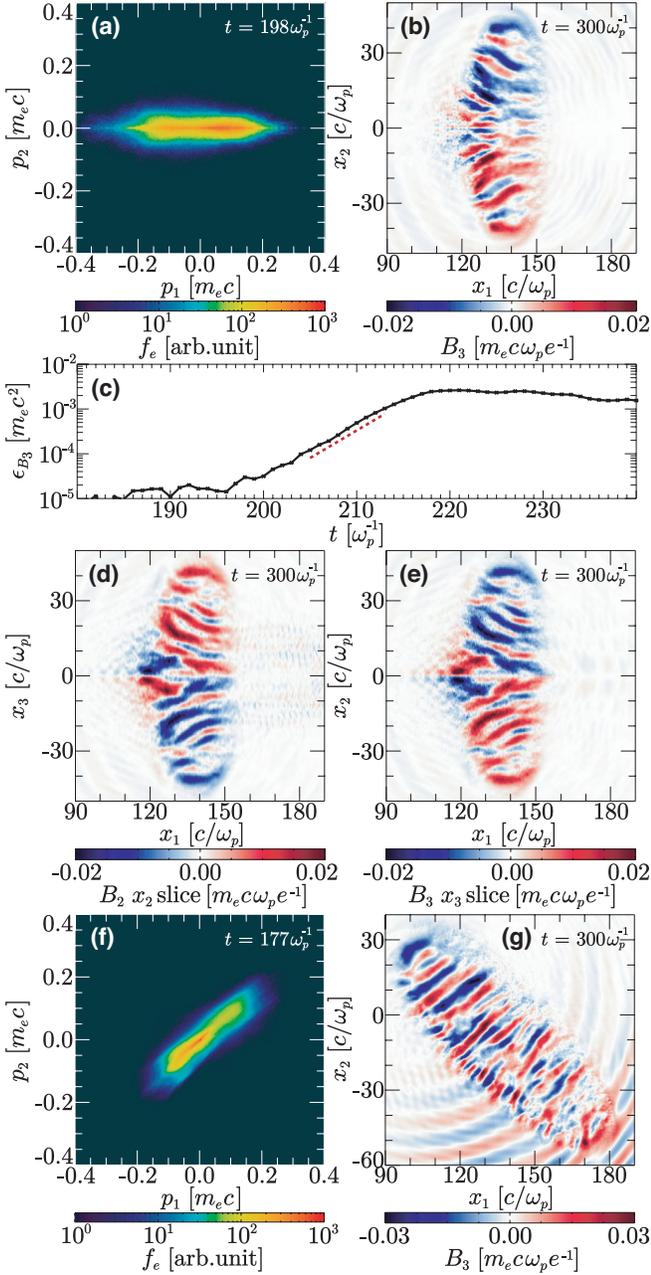}
		\caption{Seeded SRS. (a) Electron momentum distribution for seeded backscatter. (b) Magnetic field in the region where the scattered plasma waves heated the plasma. (c) $B_3$ energy and the theoretical prediction for the growth rate (dashed red line). (d-e) Slices of the magnetic field for three-dimensional simulation of seeded backscatter, the $B_2$ component at $x_2 = 0$, and the $B_3$ component at $x_3 = 0$. Seeded side-scatter (f) electron momentum distribution and (g) magnetic field in the region where the plasma was heated.}
		\label{seededraman}
	\end{figure}
	
	Figure \ref{seededraman}(a) shows the momentum space for the particles in the region where SRS grows for seeded backscatter. As expected, heating is dominant in the longitudinal direction with anisotropy parameter as high as $A> 1000$ and $T_{\text{hot}}=T_{11}\approx10\text{ keV}$. The magnetic field, shown in  Fig. \ref{seededraman}(b), exhibits a wavevector in the cold $x_2$-direction, a characteristic of the Weibel instability. The maximum growth rate for the Weibel instability, in the limit $A\gg 1$, is $\gamma/\omega_p = \sqrt{T_{\text{hot}}/m_ec^2}$\cite{1973Krall}. Figure \ref{seededraman}(c) compares this value with the growth of the $B_3$ energy calculated from the simulation, showing an excellent agreement.
	
	The seeded backscatter setup is also ideal to perform three-dimensional simulations, since all electromagnetic and plasma waves of the SRS process propagate in the longitudinal direction. Thus we are able to lower the resolution in the transverse direction, rendering the simulations less demanding. Figures \ref{seededraman}(d) and (e) show two slices of the magnetic field from three-dimensional simulations and we notice a quite striking resemblance between the field filaments, both in shape and magnitude, of the two- and three-dimensional simulations [Fig. \ref{seededraman}(b) and Fig. \ref{seededraman}(d-e)], confirming that two-dimensional simulations can not only reproduce the physical mechanism behind the whole process of magnetic field amplification, but are also able to provide quantitative predictions on the generated fields.
	
	\begin{figure}[b]
		\centering
		\includegraphics[width=0.99\linewidth]{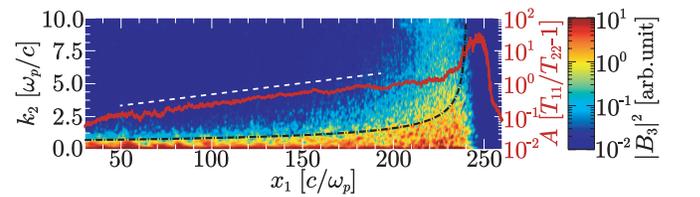}
		\caption{$\left|B_3\right|^2$ as a function of $x_1$ and $k_2$. The dot-dashed black curve fits the maximum $k_2$ of the Weibel filaments. The solid red curve is the transversely averaged anisotropy parameter $A=T_{11}/T_{22}-1$ as a function of $x_1$. The dashed white line indicates how the anisotropy decays.}
		\label{weibel_evolution}
	\end{figure}
	For seeded side-scatter, Figure \ref{seededraman}(f) shows the momentum distribution of the particles in the region where SRS grows. It is clear that the hot direction forms an angle with the $p_1$-axis. By diagonalizing the temperature tensor $T_{ij}$ [Eq. \eqref{def_temp}], we learn that the anisotropy angle is $\theta_A=40.5^\circ$. If we replace $\mathbf{k}_s = - k_s \hat{x}_2 $ in Eqs. \eqref{wv_match} and \eqref{scdisp}, we calculate that the angle of the scattered plasma waves is $\theta_{pw} = \arctan(1- \omega_p/ck_0)$, or $\theta_{pw} = 42^\circ$ for $\omega_0/\omega_p = 10$. Therefore, there is a  good agreement between the anisotropy direction and the direction of the scattered plasma waves, which reinforces the correlation between the anisotropic heating and SRS. Figure \ref{seededraman}(g) shows magnetic field in the region where the pump and the seed overlap. Again, the wavevector is parallel to the cold direction which further supports the interpretation that temperature anisotropy was the driver for the magnetic field amplification through the Weibel mechanism.
	
    \section{Conceptual experimental design}
	We indicate how this setup can be used as an experimental platform for the study of the long-time evolution of Weibel filaments. This is demonstrated on Fig. \ref{weibel_evolution} where we show the quantity $\left|B_3\right|^2$ at the same time as in Fig. \ref{fields}(c), but we made a Fourier transform in the $x_2$ direction. There is a correlation between the time delay of the laser passage through a region and the stage in the evolution of the Weibel filaments; going back in space is equivalent to go forward in time as we notice the transition of a wide range of large $k_2$ modes that grow when the anisotropy is maximum (right behind the pump) to small $k_2$ as the filaments merge. The dot-dashed black line in Fig. \ref{weibel_evolution} shows a fit $k_2\propto (c_1-x_1)^{-1/2}$ [$c_1$ is a constant] for the wavenumber trend of Weibel, in good agreement with simulation results. The wavenumber trend $\left<k(t)\right>\propto t^{-1/2}$ was also observed elsewhere \cite{2003Silva,2004Romanov,SM}. Figure \ref{weibel_evolution} shows that the decrease in wavenumber corresponds to a decrease in anisotropy (red line). The white dashed line indicates the trend $A\propto e^{-0.015\omega_{p}t}$ for long times.

	This process can be probed using technology readily available in the laboratory. For a laser wavelength of $\lambda = 800~\text{nm}$, the plasma density for this setup is $n = 1.75\times 10^{19}~\text{cm}^{-3}$ . The typical transverse dimension and intensity of the $B_3$ filaments in our simulations are $50~\mu\text{m}$ and $0.2~\text{MG}$, respectively, for a laser with $\tau=270~\text{fs}$ and $5.5~\text{TW}$ peak power. Although our fiducial simulation is collisionless considering a Hydrogen plasma (if $T>4~\text{eV}$, the collision time is greater than the anisotropy growth time), for smaller temperatures the laser will rapidly heat the plasma to a collisionless regime. Simulations with an initially cold neutral Hydrogen gas and verified that the pump fully ionizes a radius beyond the beam waist with $T\ge5~\text{eV}$ \cite{SM}. Also, the effect of a density gradient on heating due to SRS is explored in a simulation with longitudinal and transverse density gradients of $0.5\times10^{17}~\text{cm}^{-3}/\mu\text{m}$ demonstrating the robustness of our main results \cite{SM}. 
	
	\section{Summary}
	In this work, we have presented a novel mechanism for magnetic field generation in intense laser-plasma interactions. We identify magnetic field amplification that matches the Weibel instability characteristics and growth rate in the region behind the pump electromagnetic pulse. We show that Weibel is driven by anisotropic heating produced by SRS in both seeded and unseeded scenarios. We reveal that the Weibel field amplification is present as long as the electron temperature is sufficiently small that some side-scattered waves are not subject to Landau damping. Finally, this scenario enables experimental studies of the long-time evolution of the Weibel filaments, with laser parameters readily available in several facilities around the world.

	\begin{acknowledgments}
	The authors gratefully acknowledge discussions with Prof. Warren Mori and Dr. Bedros Afeyan. We would like to thank the anonymous referees for the insightful comments. This work was supported by the European Research Council through the \mbox{InPairs} project Grant Agreement No. 695088, by the EU (EUPRAXIA Grant Agreement No. 653782), and FCT (Portugal) grants PTDC/FIS-PLA/2940/2014 and SFRH/IF/01635/2015. We acknowledge PRACE for awarding us access to MareNostrum at Barcelona Supercomputing Center (BSC), Spain and SuperMUC at GCS@LRZ, Germany.
	\end{acknowledgments}

\providecommand{\noopsort}[1]{}\providecommand{\singleletter}[1]{#1}%

\end{document}